%% file: PRL_revision_wlim.tex
\documentclass[aps,prl,twocolumn,showpacs,superscriptaddress,groupedaddress,longbibliography]{revtex4-1}  
\usepackage{graphicx}  
\usepackage{dcolumn}   
\usepackage{bm}        
\usepackage{amssymb}   
\usepackage{amsmath}
\usepackage{amscd,float,array}
\usepackage{color}
\usepackage{microtype}

\newcommand{\Ree}{{\rm Re}}
\newcommand{\Ha}{{\rm Ha}}
\newcommand{\Rm}{{\rm Rm}}
\newcommand{\Pm}{{\rm Pm}}

\newcommand{\ez}{{\bf\hat e_z}}
\newcommand{\ve}{{\mathbf{v}}}

\newcommand{\lp}{\ensuremath{\left(}}
\newcommand{\rp}{\ensuremath{\right)}}

\definecolor{applegreen}{rgb}{0.55, 0.71, 0.0}
\definecolor{blue-violet}{rgb}{0.54, 0.17, 0.89}

\begin{document}


\title{Four-frequency solution in a magnetohydrodynamic Couette flow
  as a consequence of azimuthal symmetry breaking}

\input author_list.tex

\date{\today}

\begin{abstract}
The occurrence of magnetohydrodynamic (MHD) quasiperiodic flows with
four fundamental frequencies in differentially rotating spherical
geometry is understood in terms of a sequence of bifurcations breaking
the azimuthal symmetry of the flow as the applied magnetic field
strength is varied. These flows originate from unstable periodic and
quasiperiodic states with broken equatorial symmetry but having
four-fold azimuthal symmetry. A posterior bifurcation gives rise to
two-fold symmetric quasiperiodic states, with three fundamental
frequencies, and a further bifurcation to a four-frequency
quasiperiodic state which has lost all the spatial symmetries. This
bifurcation scenario may be favoured when differential rotation is
increased and periodic flows with $m$-fold azimuthal symmetry, $m$
being product of several prime numbers, emerge at sufficiently large
magnetic field.
\end{abstract}

\maketitle

Understanding how systems become chaotic is of fundamental importance
in many applications. Biological systems~\cite{Erm85,Gri06}, financial
models~\cite{Lor87}, road traffic modelling~\cite{STSAH02}, laser
physics~\cite{HMHS94}, neural networks~\cite{AlSp06}, and simulations
in fluid dynamics~\cite{CNH16}, or magnetohydrodynamics~\cite{GSGS20},
exhibit transitions from regular oscillatory behaviour to a chaotic
regime. Quite often, this transition follows the
Newhouse-Ruelle-Takens (NRT)~\cite{NRT78} scenario in which after a
few bifurcations, involving quasiperiodic states, chaos emerges.
According to the NRT theorem quasiperiodic oscillatory motions, which
are known as tori, with 3 or more fundamental frequencies are unstable
to small perturbations and thus unlikely to occur. However, the
numerical experiments of~\cite{GOY83} evidenced that the mathematical
notion of small perturbations is a key issue and that in case of an
appropriate spatial structure of the perturbations a three-tori
solution may well be observed in real nonlinear systems. Since then,
the existence of three-tori has been confirmed in experiments on
electronic circuits~\cite{CuLi88,SPM06}, solid
mechanics~\cite{AlRe00}, hydrodynamics~\cite{LPA04},
Rayleigh-B{\'e}nard convection~\cite{WKPS84} and magnetohydrodynamics
(MHD)~\cite{LFL83}. The study of MHD flows is of fundamental relevance
in geophysics and astrophysics which motivated
experiments~\cite{SGGRSSH06,GLPGS02,Gal_et_al12,Ste_et_al15} and
simulations~\cite{RoGl00,Jon11} that investigate the role of chaotic
and/or turbulent flows for planetary and stellar
dynamos~\cite{RoSt71}, or the turbulent transport processes occurring
in accretion disks~\cite{BaHa98} where the magnetorotational
instability (MRI)~\cite{BaHa91} plays a basic role.

Symmetries in physical systems provide a way to circumvent the NRT
theorem because the bifurcations occurring in these systems may be
non-generic. Their understanding is of relevance as the character of an
underlying symmetry group in general is reflected in the possible
solutions and their evolution in time, e.g.\ in terms of conservation
laws. This is especially the case in fluid dynamics~\cite{CrKn91}, or
magnetohydrodynamics, where the appearance of three-tori solutions has
been interpreted as a consequence of bifurcations~\cite{LoMa00,LFL83},
which may introduce a breaking of
symmetry~\cite{ADML12,GNS16,GSGS20}. Quasiperiodic tori with more than
three frequencies are however rarely found in systems with moderate
and large number of degrees of freedom (e.g.~\cite{AlSp06}). For
instance, the existence of four-tori solutions has been attributed to
the non-generic character of two-dimensional Rayleigh-B{\'e}nard
convection~\cite{ZSF98}, or to spatial localisation of weakly coupled
individual modes in~\cite{WKPS84,He05}. The latter studies pointed out
the relevance of the spatial structure of the solutions for the
emergence of chaos in large-scale systems.

In this Letter we investigate the emergence of four-tori and chaotic
flows in simulations of a magnetised spherical Couette (MSC)
system. Using an accurate frequency analysis based on Laskar's
algorithm~\cite{Las93} and Poincar\'e sections, we find that
consecutive symmetry breaking caused by various Hopf bifurcations
determine the evolution of the system and accompany the route to
chaotic behaviour. The MSC system constitutes a paradigmatic MHD
problem~\cite{Hol09,TEO11,FSNS13,Wic14} that is of relevance for
differentially rotating, electrically conducting flows. Flows driven
by differential rotation, which have been investigated in several
experiments~\cite{SMTHDHAL04,ZTL11,HHE16,KKSS17,KNS18,BTHW18}, govern
the dynamics in the interior of stars and/or planets~\cite{Jon11}
where they constitute a possible source for MHD wave
phenomena~\cite{Spr02}, dynamo action~\footnote{A differentially
  rotating flow, for example, provides a natural explanation for the
  strong axisymmetric character of Saturn's magnetic
  field~\cite{WiOl10}}, and perhaps even for the generation of
gravitational wave signals from neutron stars~\cite{PMGO06,Las15}.

In terms of symmetry theory~\cite{GoSt03} the MSC problem is a
{\bf{SO}}$(2)\times${\bf{Z}}$_2$ equivariant system, i.e., invariant
by azimuthal rotations ({\bf{SO}}$(2)$) and reflections with respect
to the equatorial plane ({\bf{Z}}$_2$). In {\bf{SO}}$(2)$ symmetric
systems, branches of rotating waves (RWs), either stable or unstable,
appear after the axisymmetric base state becomes unstable (primary
Hopf bifurcation~\cite{CrKn91}). Successive Hopf
bifurcations~\cite{Ran82,GLM00} give rise to quasiperiodic modulated
rotating waves (MRWs) and to chaotic turbulent flows, usually
following the NRT scenario~\cite{Ran82}. In the particular case of the
MSC, when the magnetic field is varied, branches of RWs with a
$m$-fold azimuthal symmetry with a prime number
$m=2,3$~\cite{TEO11,GaSt18} give rise to stable two- and three-tori
MRWs~\cite{GSGS19}, and eventually chaotic flows~\cite{GSGS20}, though
four-tori MRWs have not yet been found. We show below that MHD
four-tori solutions in terms of MRWs can be obtained after successive
azimuthal symmetry breaking Hopf bifurcations from a parent branch of
RW having $m$-fold symmetry which is not a prime number, in this case
$m=4$.  Note that when $m$ is a prime number only one symmetry
breaking bifurcation can take place as the flows are equatorially
asymmetric so that the case $m=4$ constitutes the lowest non-trivial
possibility for azimuthal symmetry breaking with multiple
bifurcations.

\begin{figure}
\hspace{0.mm}\includegraphics[width=0.95\linewidth]{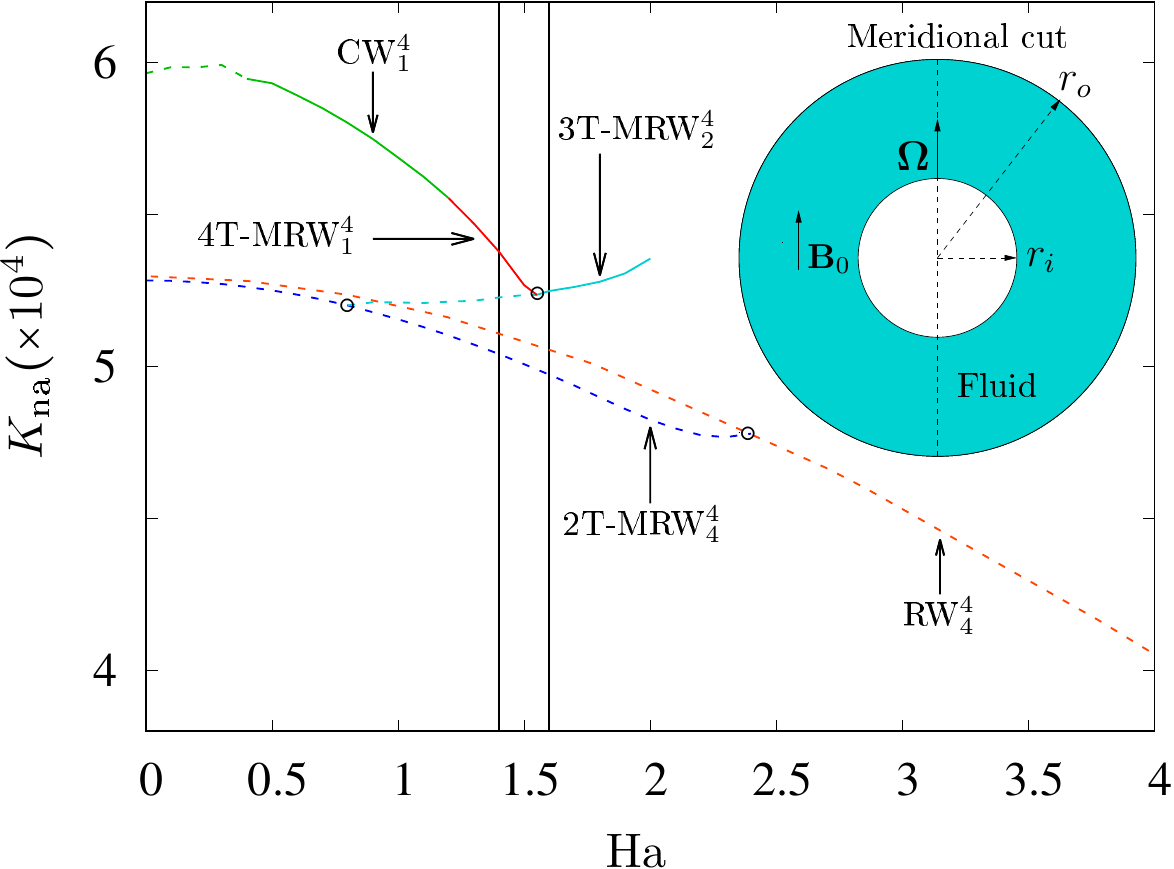}        
\caption{Magnetised spherical Couette (MSC) geometry and bifurcation
  diagram of the volume-averaged nonaxisymmetric kinetic energy
  density $K_{\text{na}}$ versus $\Ha$. Solid (dashed) lines are used
  for stable (unstable) flows. Branches of rotating waves
  RW$^{m_{\text{max}}}_m$, modulated rotating waves
  MRW$^{m_{\text{max}}}_m$, and complex waves CW$^{m_{\text{max}}}_m$
  are shown. MRWs with 2, 3 and 4 frequencies are labelled as 2T, 3T
  and 4T, respectively. The colours distinguish the various solutions
  according to the corresponding label. The vertical lines mark $\Ha$
  for the solutions selected for Fig.~\ref{fig:cont_plot}.}
\label{fig:bif}      
\end{figure}

We consider an electrically conducting fluid of density $\rho$,
kinematic viscosity $\nu$, magnetic diffusivity $\eta=1/(\sigma\mu_0)$
(where $\mu_0$ is the magnetic permeability of the free-space and
$\sigma$ is the electrical conductivity). The fluid is bounded by two
spheres with radius $r_{\text{i}}$ and $r_{\text{o}}$, respectively,
with the outer sphere being at rest and the inner sphere rotating at
angular velocity $\Omega$ around the vertical axis $\ez$ (see inset of
Fig.~\ref{fig:bif}). A uniform axial magnetic field of amplitude $B_0$
is imposed as in the HEDGEHOG experiment~\cite{KKSS17}.  Scaling the
length, time, velocity and magnetic field with
$d=r_{\text{o}}-r_{\text{i}}$, $d^2/\nu$, $r_{\text{i}}\Omega$ and
$B_0$, respectively, the temporal evolution of the system is governed
by the Navier-Stokes equation and the induction equation which read:
\begin{align}
& \partial_t\ve+\Ree\lp\ve\cdot\nabla\rp\ve = -\nabla
  p+\nabla^2\ve+\Ha^2(\nabla\times {\bf b})\times\ez , \nonumber \\ &
  0 = \nabla\times(\ve\times\ez)+\nabla^2{\bf b}, \quad
  \nabla\cdot\ve=0, \quad \nabla\cdot{\bf b}=0,\nonumber
\end{align}
where $\Ree=\Omega r_{\text{i}} d/\nu$ is the commonly known Reynolds
number, $\Ha=B_0d(\sigma/(\rho\nu))^{1/2}$ is the Hartmann number,
$\ve$ the velocity field and ${\bf b}$ the deviation of magnetic field
from the axial applied field. Here we use the inductionless
approximation which is valid in the limit of small magnetic Reynolds
number, $\Rm=\Omega r_{\text{i}} d/\eta \ll 1$. This condition is well
met when considering the liquid metal GaInSn, with magnetic Prandtl
number $\Pm=\nu/\eta\sim O(10^{-6})$~\cite{PSEGN14}, at moderate
$\Ree=10^3$ (similar to the experiment~\cite{KKSS17}) since
$\Rm=\Pm\hspace{0.5mm}\Ree \sim 10^{-3}$. The aspect ratio is
$\chi=r_{\text{i}}/r_{\text{o}}=0.5$ and no-slip
($v_r=v_\theta=v_\varphi=0$) at $r=r_{\text{o}}$ and constant rotation
($v_r=v_\theta=0,~v_\varphi= \sin{\theta}$, $\theta$ being colatitude)
at $r=r_{\text{i}}$ are the boundary conditions imposed on the
velocity field. For the magnetic field, insulating boundary conditions
are considered in accordance with typical experimental
setups~\cite{SMTHDHAL04,OGGSS20}.  Spectral methods --spherical
harmonics in the angular coordinates and a collocation method in the
radial direction-- and high order implicit-explicit
backward-differentiation (IMEX--BDF) time schemes are employed for
solving the MSC equations (see~\cite{GaSt18,GSGS20} for details).

\begin{figure*}
  \hspace{0.mm}\includegraphics[width=0.9\linewidth]{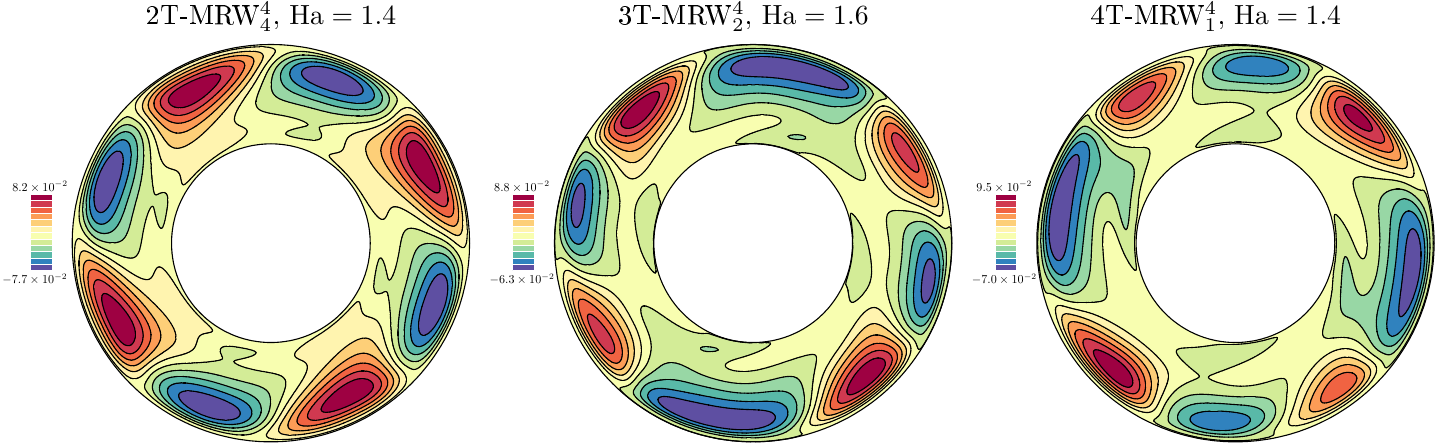}
\caption{Contour plots of the nonaxisymmetric component of the radial
  velocity on a colatitudinal section at $\theta\approx
  93^{\circ}$. The azimuthal symmetry is broken from left to right:
  from $m=4$ to $m=2$ and from $m=2$ to $m=1$ (compare the shape
    of the opposite red cells).}
\label{fig:cont_plot}      
\end{figure*}

The solutions are classified according to their azimuthal symmetry
$m$, the wave number with the largest volume-averaged kinetic energy
$m_{\text{max}}$, and their type of time dependence. In this way,
branches of RWs and MRWs are labelled as RW$^{m_{\text{max}}}_m$ and
MRW$^{m_{\text{max}}}_m$. The latter can be quasiperiodic with 2, 3
and 4 frequencies and are labelled as 2T, 3T and 4T, respectively.
The branches of equatorially asymmetric RW$^2_2$, RW$^3_3$, and
RW$^4_4$ which bifurcate from the base state at
$\Ha=12.2$~\cite{TEO11} were computed in~\cite{GaSt18} by means of
continuation methods~\cite{DoTu00,DWCDDEGHLSPSSTT14,SaNe16}.  The
latter allows, for each parameter, to find a periodic solution by
applying a Newton method to a function derived from the flow
periodicity condition (see~\cite{SaNe16} for a detailed description).
Here we focus on the analysis of MRW bifurcating from the unstable
branch RW$^4_4$ for a small control parameter $\Ha<2.5$ and fixed
$\Ree=10^3$. These MRWs have been successively obtained by means of
direct numerical simulations (DNS) of the MSC equations with $n_r=40$
radial collocation points and a spherical harmonic truncation
parameter of $L_{\text{max}}=84$. The dimension of the system is then
$n=(2L_{\text{max}}^2+4L_{\text{max}})(n_r -1)=563472$. The results
for the four-tori solution at $\Ha=1.4$ are confirmed for increased
resolution with $n_r=60$ and $L_{\text{max}}=126$. Azimuthal symmetry
$m=m_d$ can be imposed on the DNS by only considering the azimuthal
wave numbers $m=km_d,~k\in\mathbb{Z}$ in the spherical harmonic
expansion of the fields. All DNS comprise more than 100 viscous time
units and initial transients less than $10$ time units are required
before the statistically saturated state is reached. The time and
volume-averaged nonaxisymmetric kinetic energy density $K_{\text{na}}$
is employed as a proxy of the time dependence of the flows because
they initially bifurcate from an axisymmetric base state (only the
$m=0$ mode is nonzero). For each $\Ha$ a new MRW is computed from a
previous state with nearby $\Ha$. We usually use $\Delta \Ha=0.1$, but
smaller values are selected close to a bifurcation. The first branch
of MRW which we compute here, bifurcates from the unstable branch
RW$^4_4$, already computed in~\cite{GaSt18}.

\begin{figure}
\hspace{0.mm}\includegraphics[width=0.9\linewidth]{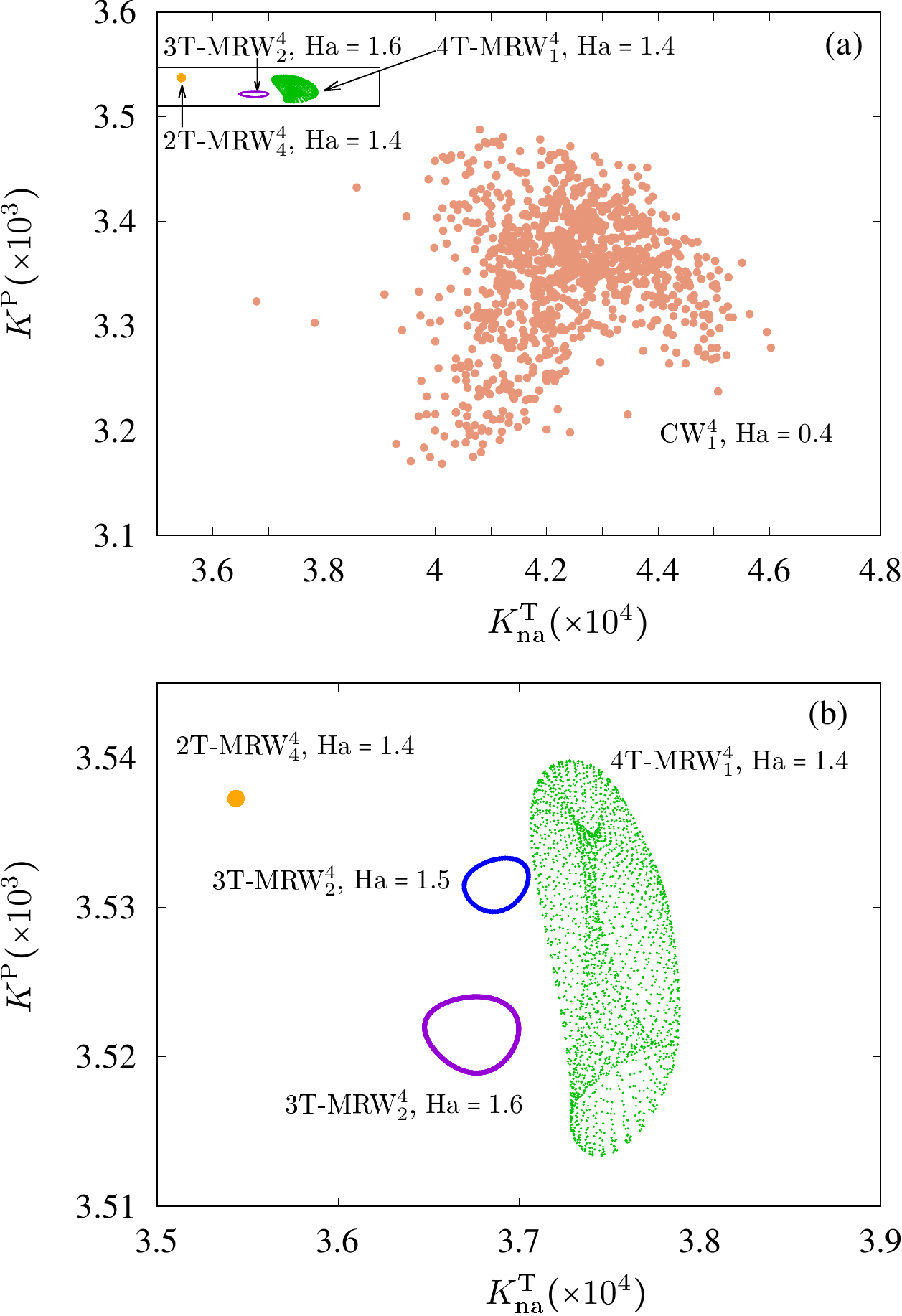}
\caption{Poincar\'e sections at the time instants $t_i$ defined by the
  constraint $K(t_i)=\overline{K}$, $K$ being the volume-averaged
  kinetic energy and $\overline{K}$ its time average. The
  volume-averaged poloidal kinetic energy $K^{\text{P}}(t_i)$ is
  displayed versus the volume-averaged toroidal nonaxisymmetric energy
  $K^{\text{T}}_{\text{na}}(t_i)$. (a) 2T-MRW$^4_4$ at $\Ha=1.4$,
  3T-MRW$^4_2$ at $\Ha=1.6$, 4T-MRW$^4_1$ at $\Ha=1.4$, and a chaotic
  wave CW$^4_1$ at $\Ha=1.4$. (b) Detail of (a) showing also a
  3T-MRW$^4_2$ at $\Ha=1.5$.}
\label{fig:poinc}      
\end{figure}

The bifurcation diagram in Fig.~\ref{fig:bif} displays $K_{\text{na}}$
versus $\Ha$ and the bifurcation points are marked with circles. By
decreasing $\Ha$, periodic RW$^4_4$ undergo a Hopf bifurcation to
2T-MRW$^4_4$ around $\Ha\approx 2.4$ which then extends down to
$\Ha=0$. These flows are obtained with time integrations with the
azimuthal symmetry constrained to $m=4$ and are unstable to small
random perturbations with azimuthal symmetry $m=1$.  Another branch of
unstable solutions appears at $\Ha=0.8$ via a secondary subcritical
Hopf bifurcation on the 2T-MRW$^4_4$ branch (blue dashed curve) which
breaks the $m=4$ symmetry to $m=2$ and finally leads to the emergence
of an unstable 3T-MRW$^4_2$ branch (light blue dashed curve).  The
latter extends for increasing $\Ha$ and becomes stable at $\Ha\approx
1.52$ where a branch of 4T-MRW$^4_1$ is born thanks to a tertiary
subcritical Hopf bifurcation breaking the $m = 2$ symmetry (red
curve). This branch is lost for $\Ha\lesssim 1.35$ (green curve) and
complex flows, either with more than four frequencies or chaotic,
occur.  The contour plots of the $m\ne 0$ component of the radial
velocity, on a colatitudinal section slightly below the equatorial
plane, are displayed in Fig.~\ref{fig:cont_plot} for one example of
each type of MRW with azimuthal symmetry $m=4, m=2$, and $m=1$ (from
left to right).

The time dependence of RWs is described by a uniform azimuthal
rotation of a fixed flow pattern whereas for MRWs the pattern is
azimuthally rotating but modulated with additional frequencies
(e.g.~\cite{Ran82}). Thus, a frequency analysis of any azimuthally
averaged property provides one frequency less than the analysis of a
localised particular flow component. Because of this, Poincar\'e
sections at the time instants $t_i$ defined by the constraint
$K(t_i)=\overline{K}$, $K$ being the volume-averaged kinetic energy
and $\overline{K}$ its time average, appear as a single point for 2T,
a closed curve for 3T, a band of points for 4T, and a cloud of points
for chaotic flows (Fig.~\ref{fig:poinc}). To confirm the regular
behavior of 2T, 3T, and 4T MRWs we perform a frequency analysis. The
frequencies giving rise to the modulation are accurately determined
from the time series of $K_4$ ($K$ restricted to the $m=4$ mode) by
means of a Fourier transform based optimisation algorithm by
Laskar~\cite{Las93}. If the solution is regular the frequencies do not
depend (within the frequency determination accuracy) on the particular
time window used for the analysis~\cite{LFC92,Las93b}. Sufficiently
wide time windows ($5-40$ time units) over large time series ($100$
time units) are considered which leads to a relative accuracy around
$10^{-5}$.

\begin{figure}
\hspace{0.mm}\includegraphics[width=0.95\linewidth]{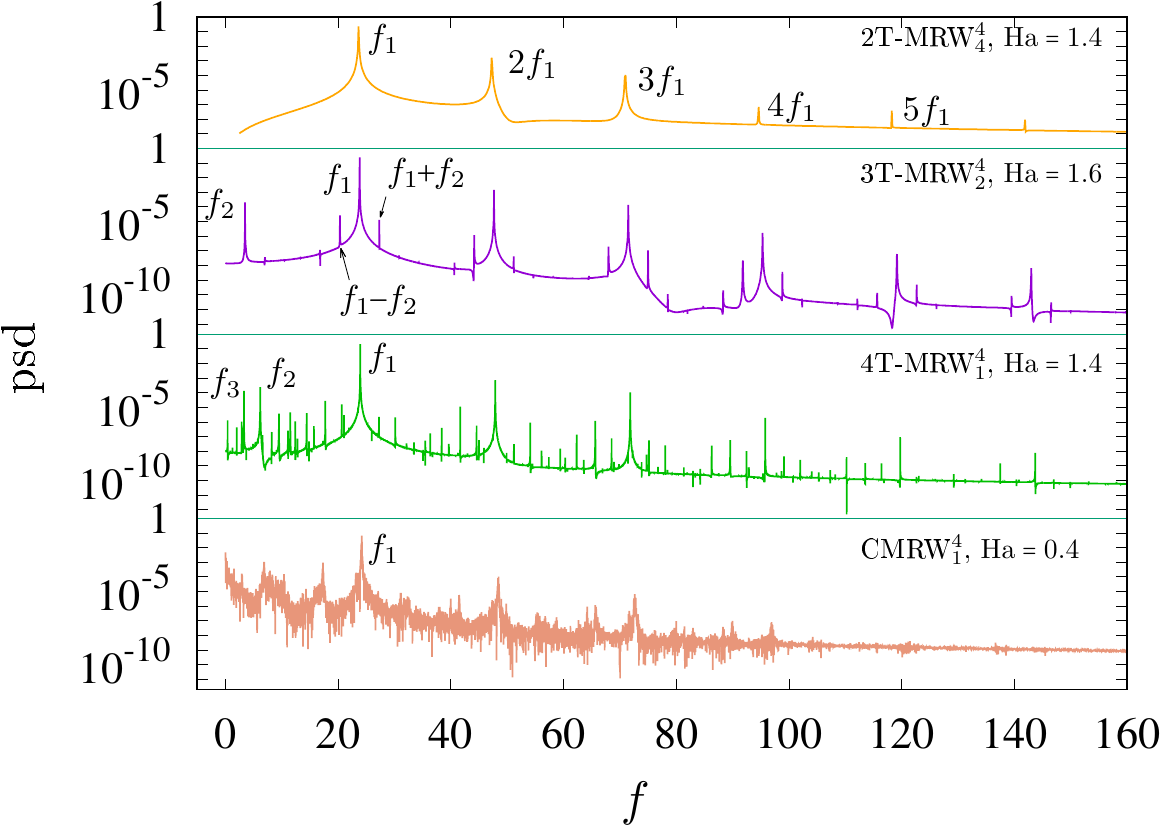}
\caption{Power spectral density for the volume-averaged kinetic
    energy density $K_4$ of the $m=4$ mode for 2T-MRW$^4_4$ at
    $\Ha=1.4$, 3T-MRW$^4_2$ at $\Ha=1.6$, 4T-MRW$^4_1$ at $\Ha=1.4$,
    and a chaotic wave CW$^4_1$ at $\Ha=1.4$. The time series of $K_4$
    are displayed on Fig.~2 of the supplementary material.}
\label{fig:tis}      
\end{figure}

The power spectral density (psd) for $K_{4}$ is displayed in
Fig.~\ref{fig:tis} for the same MRW as shown in
Fig.~\ref{fig:cont_plot} additionally including the psd for a chaotic
solution at $\Ha=0.4$ (bottom curve). Some examples for the
fundamental frequencies or corresponding linear combinations $\sum k_i
f_i$ with $k_i$ being integers, are explicitly marked.  In all cases
we have checked that the relative accuracy $|f - \sum k_i f_i|/f
<10^{-5}$, with $-6 \leq k_i \leq 6$, for all frequencies obtained
with Laskar's algorithm. In the simplest case, the 2T-MRW$_4^4$ at
$\Ha=1.4$, only one fundamental frequency $f_1$ and its multiples are
present, because the fundamental frequency associated to the drift
motion is removed by volume averaging. A second frequency $f_2$ and
few combinations $k_1 f_1 + k_2 f_2$ occur for the 3T-MRW$^4_2$
whereas the psd of 4T-MRW$^4_1$ reveals a complex time dependence with
several combinations of the type $k_1 f_1 + k_2 f_2 + k_3 f_3$. For
$\Ha \lesssim 1.35$ the variation of the main frequency $f_1$ becomes
larger than the accuracy and thus 4T-MRW$^4_1$ give rise to more
complex motions, which could be regular flows with a very small
additional frequency or chaotic flows. However, deciding whether a new
very small frequency has appeared in this regime would require
extremely long time integrations of the MSC system, which are out of
the scope of the present study. Nevertheless, it is clear that fully
broadband frequency distribution characteristic for chaotic flows is
obtained for $\Ha \lesssim 0.7$. The psd for the chaotic solution at
$\Ha=0.4$ shown in Fig.~\ref{fig:tis} (bottom curve) exhibits a
noticeable peak that corresponds to the main frequency of modulation
of the original MRW.
  
\begin{figure}
\hspace{0.mm}\includegraphics[width=0.95\linewidth]{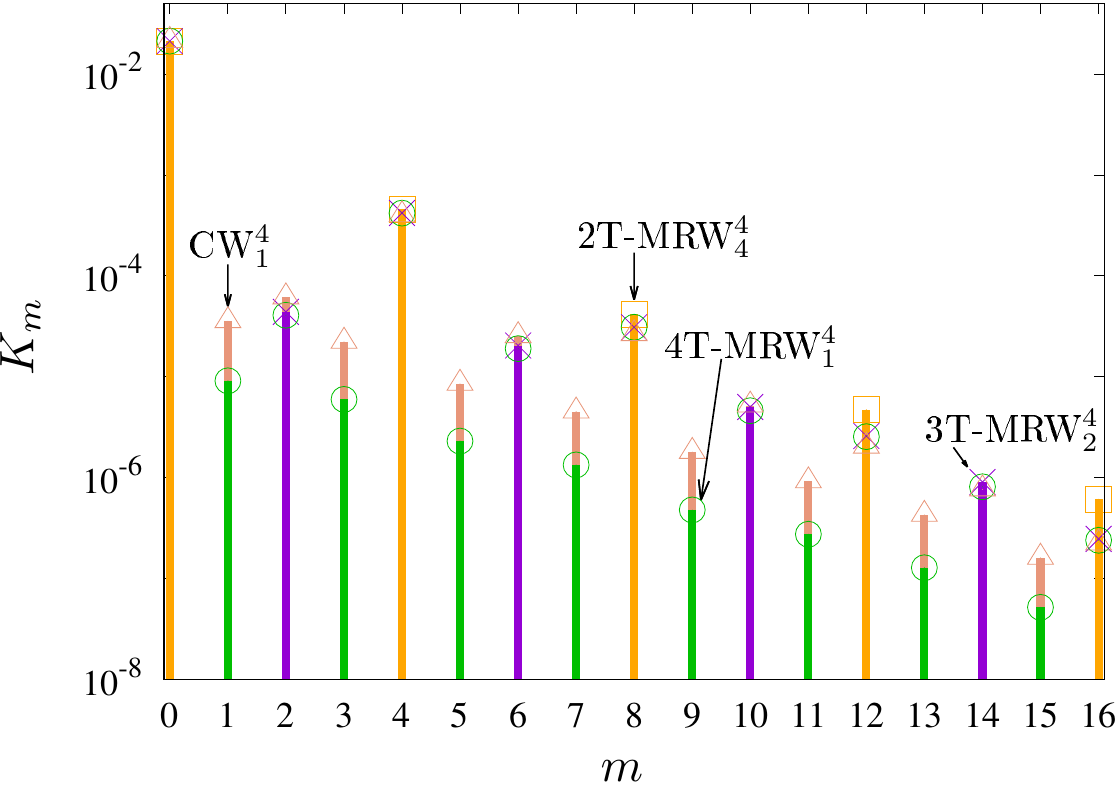}
\caption{(a) Kinetic energy spectra ($K_{m}$ versus $m$) for
  2T-MRW$^4_4$ at $\Ha=1.4$ ({\footnotesize $\square$}), 3T-MRW$^4_2$
  at $\Ha=1.6$ ({$\times$}), 4T-MRW$^4_1$ at
  $\Ha=1.4$ ({\Large $\circ$}), and a chaotic flow CW$^4_1$ at
  $\Ha=0.4$ ({$\bigtriangleup$}).}
\label{fig:ener_spec}      
\end{figure}

As suggested in~\cite{WKPS84}, the strong polar localisation of the
$m=1$ (also $m=3$) perturbations (see Fig. 1, supplementary material)
provides a way to overcome the NRT requirements and thus explains the
existence of 4T-MRW$^4_1$. In our case, successive azimuthal symmetry
breaking bifurcations from unstable regular states (that can not be
realised in the experiments of~\cite{WKPS84}) are responsible for the
mode localisation. For the regular solutions, the kinetic energy
fluctuations of the different modes evidence a weak nonlinear
interaction, being the $m=4$ component which most contributes to the
flow (see Figure 2 supplementary material, also the kinetic energy
spectra in Fig.~\ref{fig:ener_spec}). The amplitude of fluctuations
leading to chaotic flows seem not to be associated to turbulent
spatial behaviour as the $m=4$ component of the flow still dominates
over the other modes (Fig.~\ref{fig:ener_spec}).

The present study evidences that four-tori solutions are physically
possible in MHD problems and can be explained in terms of bifurcation
theory. The bifurcation scenario resembles the NRT scenario but
involves two additional Hopf bifurcations, including a subcritical
bifurcation leading to a stabilisation of a three-tori solution. This
kind of stabilisation was also found in~\cite{MaTu95}, but for
axisymmetric steady states, arising in purely hydrodynamic spherical
Couette (SC) flows. Further SC experiments~\cite{WER99} are in
accordance with the NRT scenario, but only two-tori were detected
before the regime of chaotic flows. Similarly, three or four-tori have
not been found in a comprehensive study of the different flow regimes
in the SC system with positive or negative differential
rotation~\cite{Wic14}. When the magnetic field is included, several
MSC regimes have been studied recently~\cite{Hol09,GJG11,KNS18} but
the existence of quasiperiodic flows with three frequencies has not
been shown until~\cite{GSGS20}.

The fundamental result presented here is that in a system with
symmetry, more symmetry breaking bifurcations may be required in the
NRT scenario until a flow can become chaotic, and thus regular motions
with three or four frequencies are likely to occur, formed by modes
localised in different parts of the domain~\cite{WKPS84}. These flows
originate from unstable states of azimuthal symmetry $m=4$ and thus
their origin can only be understood with symmetrically constrained
simulations and not with experiments. Speculatively, periodic flows
with an azimuthal symmetry that constitute a product of several prime
numbers, may give rise to quasiperiodic flows with more than four
frequencies, as several symmetry breaking bifurcations can occur. The
new frequency occurring at the bifurcation is usually smaller (but
also can be larger, e.\,g.~\cite{WKPS84}) than the previous
fundamental one, as can be seen in Fig.~\ref{fig:tis}. With on going
bifurcations the effect requires a longer time evolution of the system
until the upcoming frequency can be detected in the captured time
series.  The smallest fundamental frequency of the simulated
4T-MRW$^4_1$ at $\Ha=1.4$ gives rise to a time scale of around 1600
seconds which should be detectable in the HEDGEHOG
experiment~\cite{KKSS17}, whose duration is limited to 10 hours due to
the decrease of signal quality of flow measurement~\cite{OGGSS20}.
The results also bear relevance for the MRI as previous
experiments~\cite{SMTHDHAL04} may be understood in terms of MSC
instabilities~\cite{Hol09,GJG11}, similar to those analysed here.

\section{Supplementary material}

\begin{figure}[t!]
  \hspace{-2.mm}\includegraphics[width=0.8\linewidth]{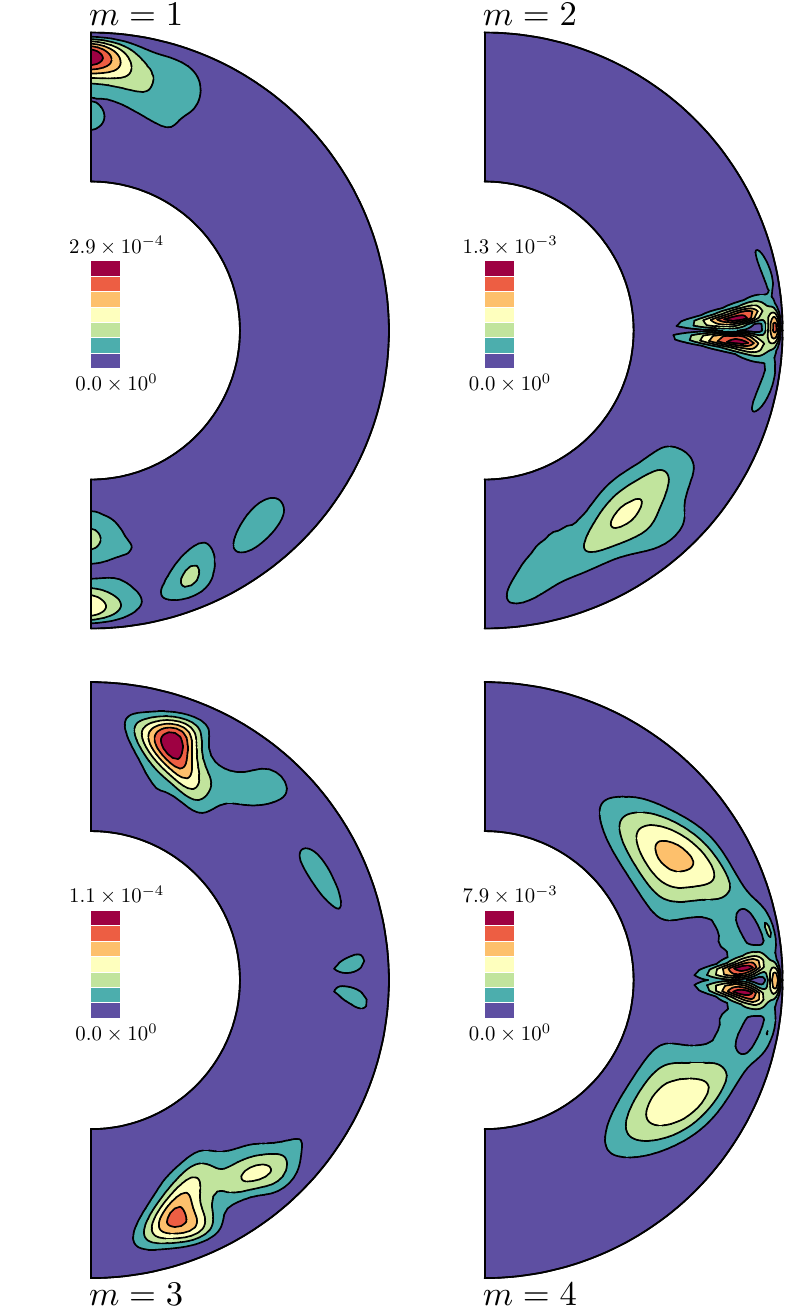}
\caption{Contour plots of the $m=1,2,3,4$ components (from left to
  right and from top to bottom) of the kinetic energy density on a
  meridional section through a relative maximum. The solution is the
  4T-MRW$^4_1$ at $\Ha=1.4$ displayed on the rightmost plot of Fig. 2
  of the main manuscript.}
\label{fig:cont_plot2}      
\end{figure}

\begin{figure}[t!]
\hspace{0.mm}\includegraphics[width=0.7\linewidth]{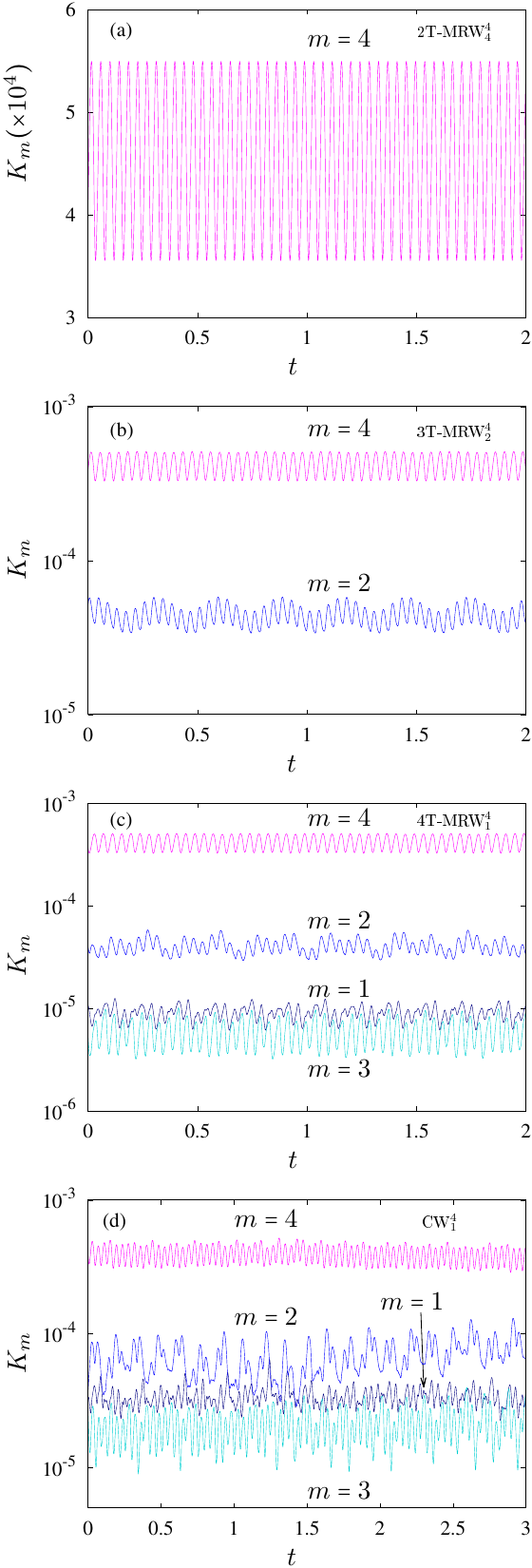}
\caption{Time series of the volume-averaged kinetic energy contained
  in each azimuthal wave number $m=1,2,3,4$. (a) Two frequency
  solution 2T-MRW$^{4}_4$ at $\Ha=1.4$. Only the modes
  $m=4k,~k\in\mathbb{Z}$ are nonzero. (b) Three frequency solution
  3T-MRW$^{4}_2$ at $\Ha=1.6$. Only the modes $m=2k,~k\in\mathbb{Z}$
  are nonzero. (c) Four frequency solution 4T-MRW$^{4}_1$ at
  $\Ha=1.4$. All the modes $m\in\mathbb{Z}$ are nonzero. (d) Chaotic
  solution at $\Ha=0.4$. Panels (a),(b) and (c) correspond to the
  contour plots, from left to right respectively, shown in figure 2 of
  the main manuscript. The power spectral density of the time series
  corresponding to $K_4$ are displayed on figure 4 of the main
  manuscript.}
\label{fig:ts_ener}      
\end{figure}

The modulated rotating waves (MRWs) presented in the main manuscript
have been successively obtained by means of direct numerical
simulations (DNS) of the magnetised spherical Couette (MSC) equations
with $n_r=40$ radial collocation points and a spherical harmonic
truncation parameter of $L_{\text{max}}=84$. The dimension of the
system is then $n=(2L_{\text{max}}^2+4L_{\text{max}})(n_r
-1)=563472$. The results for the four-tori solution at $\Ha=1.4$ are
confirmed for increased resolution with $n_r=60$ and
$L_{\text{max}}=126$. Azimuthal symmetry $m=m_d$ can be imposed on the
DNS by only considering the azimuthal wave numbers
$m=km_d,~k\in\mathbb{Z}$ in the spherical harmonic expansion of the
fields. All DNS comprise more than 100 viscous time units and initial
transients less than $10$ time units are required before the
statistically saturated state is reached.  For each $\Ha$ a new MRW is
computed from a previous state with nearby $\Ha$. We usually use
$\Delta \Ha=0.1$, but smaller values are selected close to a
bifurcation. The first branch of MRW which we compute in the present
study, bifurcates from the unstable branch RW$^4_4$, already computed
in~\cite{GaSt18}.

Figure~\ref{fig:cont_plot2} corresponds to the 4T-MRW$^4_1$ at
$\Ha=1.4$ and displays, from left to right and from top to bottom, the
contour plots of the kinetic energy density, on a meridional section
through a relative maxima, for the $m=1$, $m=2$, $m=3$ and $m=4$
modes, respectively. While fluid motions remain confined to mid and
low latitudes in the case of $m=2$ and $m=4$ modes (even modes), the
flow is restricted to high latitudes in the case of odd modes ($m=1$
and $m=3$). Then, even modes contribute to motions outside the tangent
cylinder (an imaginary cylinder parallel to the rotation axis and
tangent to the inner sphere) whereas odd modes contribute to motions
within the tangent cylinder. As suggested in~\cite{WKPS84} the
different localisation of the different modes within the domain may
explain the appearance of flows with four fundamental frequencies.

To describe the nature of kinetic energy fluctuations, the time series
of the volume-averaged kinetic energy of each azimuthal wave number
$m=1,2,3,4$ is displayed in figure~\ref{fig:ts_ener}(a-c) for the
three types of MRW shown in Figs. 2,4 and 5 of the main manuscript:
A 2T-MRW$^4_4$, a 3T-MRW$^4_2$, and a 4T-MRW$^4_1$. The dominance of
the $m=4$ component of the flow and the quasiperiodic character of the
waves are clear from the figure. The modes arising at the symmetry
breaking bifurcations ($m=1,2,3$) have larger fluctuations than the
main $m=4$ component of the flow, but barely contribute to the total
kinetic energy. The time series for a chaotic solution CW$^4_1$ is
plotted in Fig.~\ref{fig:ts_ener}(d) and shows a larger contribution
of the non-dominant $m=1,2,3$ modes.

\input acknowledgement.tex 

\end{document}

%% file: author_list.tex
\affiliation{Department of Magnetohydrodynamics, Helmholtz-Zentrum Dresden-Rossendorf, Bautzner Landstra\ss e 400, D-01328 Dresden, Germany}
\author{F.~Garcia, M.~Seilmayer, A.~Giesecke, and F.~Stefani}





%
%
%
\vskip 0.25cm

%% file: acknowledgement.tex
%
F.~G. was supported by the Alexander von Humboldt Foundation. This
project has also received funding from the European Research Council
(ERC) under the European Union’s Horizon 2020 research and innovation
programme (grant agreement No 787544).